\documentclass[aps,prl,floatfix,twocolumn,footinbib]{revtex4}
\usepackage{epsfig}
\usepackage{amsmath,amssymb}
\begin{document}

\title{Resonant tunneling magneto resistance in coupled quantum wells}

\author{Christian Ertler\footnote{email: christian.ertler@physik.uni-regensburg.de}}
\author{Jaroslav Fabian\footnote{email: jaroslav.fabian@physik.uni-regensburg.de}}
\affiliation{Institute for Theoretical Physics, University of
Regensburg, D-93040 Regensburg, Germany}

\begin{abstract}

A three barrier resonant tunneling structure in which the two
quantum wells are formed by a dilute magnetic semiconductor material
(ZnMnSe) with a giant Zeeman splitting of the conduction band is
theoretically investigated. Self-consistent numerical simulations of
the structure predict giant magnetocurrent in the resonant bias
regime as well as significant current spin polarization for a
considerable range of applied biases.

\end{abstract}
\maketitle

Semiconductor spintronics offers new functionalities to
the existing electronics technology by combining charge
and spin properties of the current carriers \cite{ZuFaSa04}.
The goal of spintronic devices is to modulate charge current
by changing the electron spin. In magnetic tunnel junctions, for example,
magnetic electrodes are sandwiched around a nonmagnetic tunnel barrier,
yielding large magnetocurrents (relative current magnitudes for parallel and
antiparallel orientations of the electrodes' magnetizations)
\cite{ZuFaSa04}.
In other spintronic devices, e.g., magnetic resonant
tunneling diodes \cite{SlGoSl03, HaTaAs00, BrWu98, VuMe03, VoShLe00, GrKeFi01, BeBeBo05},
the transmission at a given voltage can depend
strongly on the spin orientation of the electrons at the Fermi
level. Such diodes have been used as spin filters or spin detectors.

Here we propose to use magnetic resonant tunneling diodes comprising
coupled magnetic quantum wells and three nonmagnetic barriers, as
spintronic devices offering large magnetocurrents. Such devices
combine the geometry of  magnetic tunnel junctions with magnetic
resonant tunneling diodes. Corresponding three-barrier nonmagnetic
structures have been experimentally studied \cite{ZoNoOb89,
PaBeCa90, MaBrCl96, NeSoMe05}. The magnetic three barrier structure
allows to establish parallel and antiparallel magnetization
configurations and to observe magnetocurrent. The origin of the
magnetization here is not essential. The quantum wells can be either
dilute magnetic semiconductors \cite{Fu88}, which have giant
g-factors at temperatures up to 100 K, or they could be
ferromagnetic quantum wells \cite{OiMoKa04, FuWoLi04, Di02} in which
the splitting is given by the exchange field. Performing realistic
selfconsistent calculations of the I-V characteristics we indeed
predict large magnetocurrents (say, 10000\% for a moderate spin
splitting of 10 meV) at resonant voltages. These large values appear
because of the suppression of the resonant tunneling due to the
energy mismatch for the resonant levels of the coupled wells in the
antiparallel configuration. Although the predicted effect is rather
robust, we provide hints on how to further tune the heterostructure
parameters to maximize the magnetocurrent.

\begin{figure}
 \centerline{\psfig{file=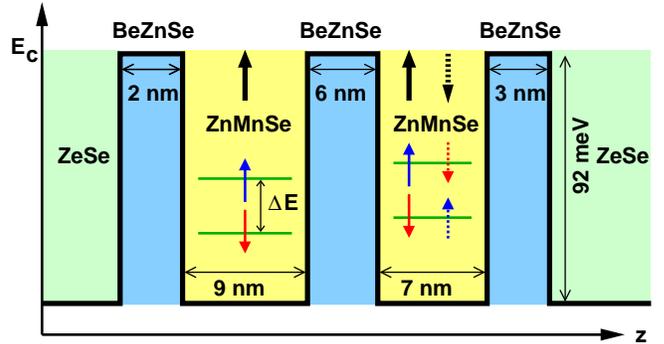,width=1\linewidth}}

\caption{(Color online) Schematic conduction band profile of an all II-IV
semiconductor three barrier heterostructure. The quasibound states
in the quantum wells are Zeeman-split by an external magnetic
field.} \label{fig:scheme}
\end{figure}

We investigate a three barrier semiconductor
heterostructure based on a Zn(Be,Mn)Se material system. The
conduction band profile at zero bias is schematically shown in
Fig. \ref{fig:scheme}. The barriers are formed by Be-doped ZnSe,
whereas the quantum wells (QWs) are made of the
dilute magnetic semiconductor (DMS) material ZnMnSe. Such a system is an
extension of a two barrier structure already experimentally investigated in Ref. \cite{SlGoSl03}.
Following a previous theoretical work \cite{HaTuVa05}
on two-barriers structures, we assume the barrier height to be 92 meV, which is about
23$\%$ of the band gap difference between ZnSe and ZnBeSe. In
DMSs an external magnetic field $B$ causes a giant Zeeman energy
splitting $\Delta E$ into spin up and spin down electron states,
which can be expressed by the modified Brillouin function $B_s$
\cite{BeBeBo05}:
\begin{equation}
\Delta E  = x_{\mathrm{eff}} N_0\alpha s B_s(g_{Mn} s \mu_B B/k_B
T_{\mathrm{eff}}).
\end{equation}
Here, $x_{\mathrm{eff}}$ is the effective concentration of Mn ions,
$N_0\alpha= 0.26$ eV is the $sp-d$ exchange constant for conduction
electrons, $s= 5/2$ is the Mn spin, $g_{Mn} =2.00$ is the Mn
g-factor, $\mu_B$ labels the Bohr magneton, $k_B$ denotes Boltzmanns
constant, and $T_{\mathrm{eff}}$ is an effective temperature. At low
temperatures (all simulations are performed at a lattice temperature
of $T$= 4.2 K), Mn concetrations of about 8$\%$, and practical magnetic fields of  1-2 T, the giant
electron g-factor in the DMS leads to an energy splitting of the
conduction band of about $10$ meV \cite{SlGoSl03}. The whole structure is considered
to be sandwiched between two leads consisting of $n$-doped ZeSe ($n
= 10^{18}$ cm$^{-3}$) of 10 nm width. Similar to experiments
\cite{SlGoSl03}, we include $5$ nm thick undoped ZnSe buffer layers
between the leads and the active structure, which causes an upward
band bending at zero bias.

Following the classic treatments of (two barriers) resonant tunneling
diodes \cite{VaLeLo83, CaMcDa87, Po89}, we assume coherent
transport throughout the whole active region. This is a reasonable assumption also for
our three barriers magnetic diode, since signatures of coherence, namely the
splitting of the well states into bonding and antibonding states, have been observed
experimentally in such structures \cite{ZoNoOb89}. We calculate the spin
dependent current flow by numerically solving the conduction band
effective mass Schr\"odinger equation taking into account the spin
dependent potential energy,
\begin{equation}
U_i(z)-e\phi(z)+ \sigma \Delta E(z).
\end{equation}
Here, $U_i$ is the intrinsic conduction band profile, $e$ is the
elementary charge and $\sigma = \pm 1/2, (\uparrow, \downarrow)$
labels the spin quantum number. The electrostatic potential $\phi$
is obtained from the Poisson equation, which is solved together with
the Schr\"odinger equation in a self-consistent way. The current
density of electrons with spin $\sigma$ is calculated by
\begin{equation}\label{eq:j}
j_\sigma = \frac{e}{(2\pi)^3}\int \mathrm{d}^3k \:v_z
T_\sigma(E_l,E_t)[f(E)-f(E+eV_a)],
\end{equation}
where $E_l$ and $E_t$ are, respectively, the longitudinal and
transverse component of the electron total energy $E$, $V_a$ denotes
the applied voltage, $T_\sigma(E_l,E_t)$ is the electron
transmission function, $v_z$ labels the longitudinal component of
the electron group velocity, and $f(E) = 1/[1+\exp(E-E_f)/k_B T]$ is
the Fermi function with the fermi energy $E_f$. Assuming parabolic
bands and using the same effective mass $m_*/m_0 = 0.145$ (with
$m_0$ denoting the free electron mass) for all layers of the
heterostructure, the transmission function only depends on the
longitudinal energy, $T(E_l,E_t) = T(E_l)$. This allows to reduce
Eq. (\ref{eq:j}) to the Tsu-Esaki formula \cite{TsEs73}. The current
spin polarization is then determined by $P_j =
(j_\uparrow-j_\downarrow)/(j_\uparrow+j_\downarrow)$.

\begin{figure}
 \centerline{\psfig{file=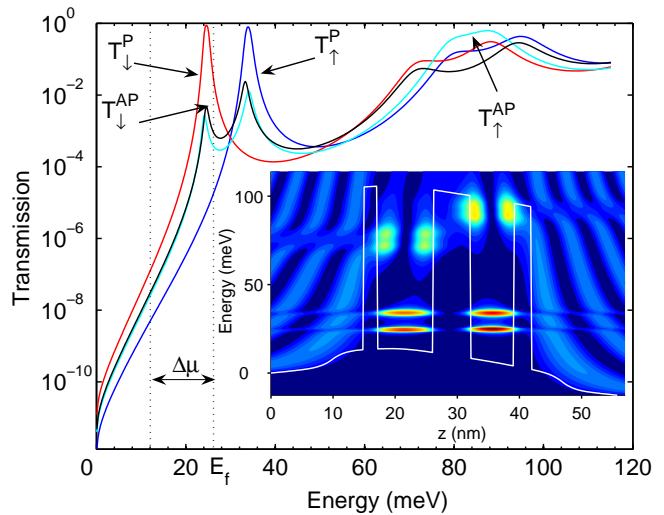,width=1\linewidth}}

\caption{(Color online) The spin resolved energy-dependent
transmission function at T = 4.2 K for parallel (P) and antiparallel
(AP) magnetization in the case of a given Zeeman splitting of
$\Delta E = 10$ meV and a resonant applied voltage of $V_a = 13$
meV. The inset shows a contour plot of the local density of states
versus energy and growth direction $z$ at resonance ($V_a = 13$ mV)
for parallel magnetization and $\Delta E = 10$ meV. The solid line
indicates the self-consistent conduction band profile. The higher
quasibound states (at about 80 and 90 meV) are not in resonance.}
\label{fig:T}
\end{figure}

We assume that the magnetization of the first QW is fixed, whereas
the second is 'soft', which means that it is sensitive to local
changes of an external magnetic field. Hence, a magnetocurrent MC
can be defined as the relative difference of the current $I$ for
parallel (P) and antiparallel (AP) alignment of the magnetization of
the two quantum wells, $MC =(I_P-I_{AP})/I_{AP}$. The proposed
structure aims in producing very high MCs based on the following
idea of operation. High resonant tunneling throughout the whole
structure is possible if the middle barrier is thin enough and if
two quasi-bound states of same spin of the adjacent QWs are aligned
energetically. Such resonant condition for the lowest energy states in the
case of P magnetization is illustrated in the inset of Fig. \ref{fig:T}. The inset shows the
local density of states of the conduction electrons and clearly
demonstrates the Zeeman splitting of the quasibound states into spin
up and spin down states. Here, the orientation of the magnetic field
is assumed such that the spin down (up) energy levels are shifted
downwards (upwards) in energy. Due to the interaction of the two QWs
the resonant energy levels are further split into bonding and
antibonding states. The splitting is observable in the local density
of states if it is greater than the natural energy broadening of the
quasibound state. For our structure the middle barrier is too thick
to resolve this additional splitting in the plot of the local
density of states. The inset of Fig. \ref{fig:T} also shows higher
quasibound states at about 80 and 90 meV, which are however not in
resonance.

In the case of P magnetization the spin up and down quasibound states
are equally Zeeman shifted in both QWs, whereas for AP alignment
they are shifted energetically in opposite directions. Assuming QWs
of the same width, i.e, with the same quasi-bound energy spectrum, there
are hence at equilibrium open resonant conduction channels for the P
alignment, whereas the transmission is blocked for the AP
configuration. However, when a small finite voltage is applied to
the structure the resonant channels of the P magnetization are
'destroyed', since the energy levels of the second QW are shifted
more deeply to lower energies than those of the first QW by the
applied bias. To overcome this shortcoming, we propose to use asymmetric
QWs. Here, we take the second QW
to be thinner than the first one. This gives rise to a higher ground
state energy in the second QW and the resonant condition is
therefore adjusted at a finite voltage, leading to high currents.

\begin{figure}
 \centerline{\psfig{file=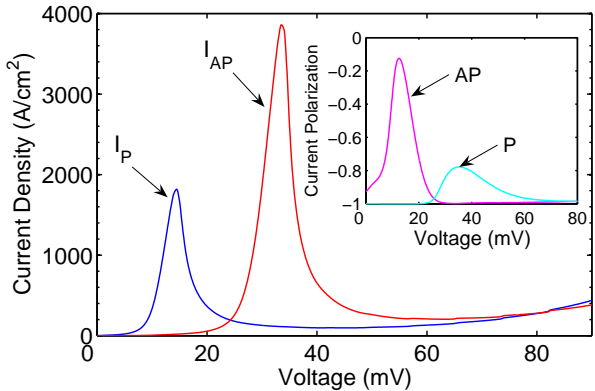,width=1\linewidth}}
\caption{(Color online) Current-Voltage characteristics of the structure for
parallel (P) and antiparallel (AP) magnetization at the temperature
T = 4.2 K and the Zeeman splitting of $\Delta E$ = 10 meV.
The inset displays the current polarization as a function
of the applied voltage.} \label{fig:IV}
\end{figure}

In order to maximize the MC we pursue the following strategy for
determining the different layer widths. At low temperatures the
current is given, in good approximation, by the quadrature of the
transmission function over the energy window $\Delta\mu =
[E_f-eV_a,E_f]$. To obtain high currents for the P alignment, the
layer widths should be chosen such, that the resonant tunneling
condition, $T(E)\approx 1$, is fulfilled at a finite voltage for
energies belonging to the interval $\Delta\mu$. On the other hand,
for the AP configuration the transmission function should be made as
small as possible in the energy window $\Delta\mu$. To meet both
demands at the same time, we use a relatively thin middle barrier,
which effectively controls the coupling of the QWs.  Figure
\ref{fig:T} shows the spin-resolved transmission function versus
energy for the resonant voltage $V_a = 13$ mV. The double peak structure of the 
transmission function for the AP configuration corresponds to the lowest quasibound state in both
QWs, which have in that case different energies leading to two 'half-resonances'. The
transmission for the AP alignment can be strongly hampered by
choosing thick QWs. The variation of the thickness of the first and
third barrier barely influences the MC, since the transmission is
then either reduced or increased for both P and AP magnetization. By
changing the buffer layer thicknesses one can tune the amount of
band bending and hence, the relative position of the quasi-bound
energies to the Fermi level. The dimensions of the structure finally
used in our numerical calculations are indicated in Fig.
\ref{fig:scheme}.

\begin{figure}
 \centerline{\psfig{file=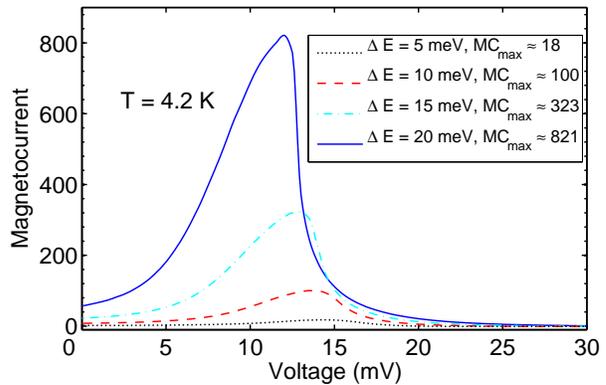,width=1\linewidth}}
\caption{(Color online) Magnetocurrent (MC) as a function of applied
voltage for different Zeeman splittings $\Delta E$.} \label{fig:MC}
\end{figure}

The obtained current-voltage characteristics for P and AP
magnetization and a fixed Zeeman splitting of $\Delta E = 10$ meV
are displayed in Fig. \ref{fig:IV}. In the case of P magnetization,
the lowest quasibound energy levels of the adjacent QWs are already
aligned at a small voltage of about $13$ mV, whereas for the AP
configuration a much higher voltage is necessary to obtain resonant
tunneling. The current spin polarization is plotted in the inset of Fig.\ref{fig:IV}.
Since the spin down state has a lower energy than the
spin up state, the current for P magnetization is almost all spin down polarized at 
low voltages. For higher voltages also spin up states contribute to the current thereby diminishing
the polarization. In the case of AP magnetization, spin up current can flow at low voltages due to the
lowest spin up state in the second QW. By inreasing the voltage this state gets off resonance, which 
leads again to significant spin down polarization.  
The obtained MCs for different Zeeman
energy splittings are shown in Fig \ref{fig:MC}. The transmission for AP
magnetization is strongly reduced by increasing the energy splitting
$\Delta E$. Hence, the current for AP magnetization becomes very
small at the peak voltage of the P configuration and our simulations
reveal very high MCs up to 800 for reasonable spin splitting.
Simulations performed at a higher temperature $T=100$ K show that the MC is reduced to about 25$\%$ of 
its value at $T=4.2$ K. Although the temperature effect is quite large, the MC remains significant.  

To summarize, we have numerically investigated an 
all semiconductor three barrier resonant tunneling structure,
comprising two quantum wells made of a magnetic material. Our simulations predict very high
magnetocurrents demonstrating the device potential of the proposed structure.
Further investigations of an {\em in-plane} (perpendicular to the growth 
direction of the heterostructure) resistance of the two coupled magnetic QWs
promise to reveal interesting effects, since resistance resonances for such a current-in-plane 
configuration have already been demonstrated for 
nonmagnetic structures  \cite{BeKaPa94}.

This work has been supported by the Deutsche Forschungsgesellschaft SFB 689.





\begin{thebibliography}{22}
\expandafter\ifx\csname natexlab\endcsname\relax\def\natexlab#1{#1}\fi
\expandafter\ifx\csname bibnamefont\endcsname\relax
  \def\bibnamefont#1{#1}\fi
\expandafter\ifx\csname bibfnamefont\endcsname\relax
  \def\bibfnamefont#1{#1}\fi
\expandafter\ifx\csname citenamefont\endcsname\relax
  \def\citenamefont#1{#1}\fi
\expandafter\ifx\csname url\endcsname\relax
  \def\url#1{\texttt{#1}}\fi
\expandafter\ifx\csname urlprefix\endcsname\relax\def\urlprefix{URL }\fi
\providecommand{\bibinfo}[2]{#2}
\providecommand{\eprint}[2][]{\url{#2}}

\bibitem[{\citenamefont{\v{Z}uti\'{c} et~al.}(2004)\citenamefont{\v{Z}uti\'{c},
  Fabian, and {Das Sarma}}}]{ZuFaSa04}
\bibinfo{author}{\bibfnamefont{I.}~\bibnamefont{\v{Z}uti\'{c}}},
  \bibinfo{author}{\bibfnamefont{J.}~\bibnamefont{Fabian}}, \bibnamefont{and}
  \bibinfo{author}{\bibfnamefont{S.}~\bibnamefont{{Das Sarma}}},
  \bibinfo{journal}{Rev. Mod. Phys.} \textbf{\bibinfo{volume}{76}},
  \bibinfo{pages}{323} (\bibinfo{year}{2004}).

\bibitem[{\citenamefont{Slobodskyy et~al.}(2003)\citenamefont{Slobodskyy,
  Gould, Slobodskyy, Becker, Schmidt, and Molenkamp}}]{SlGoSl03}
\bibinfo{author}{\bibfnamefont{A.}~\bibnamefont{Slobodskyy}},
  \bibinfo{author}{\bibfnamefont{C.}~\bibnamefont{Gould}},
  \bibinfo{author}{\bibfnamefont{T.}~\bibnamefont{Slobodskyy}},
  \bibinfo{author}{\bibfnamefont{C.~R.} \bibnamefont{Becker}},
  \bibinfo{author}{\bibfnamefont{G.}~\bibnamefont{Schmidt}}, \bibnamefont{and}
  \bibinfo{author}{\bibfnamefont{L.~W.} \bibnamefont{Molenkamp}},
  \bibinfo{journal}{Phys. Rev. Lett.} \textbf{\bibinfo{volume}{90}},
  \bibinfo{pages}{246601} (\bibinfo{year}{2003}).

\bibitem[{\citenamefont{Hayashi et~al.}(2000)\citenamefont{Hayashi, Tanaka, and
  Asamitsu}}]{HaTaAs00}
\bibinfo{author}{\bibfnamefont{T.}~\bibnamefont{Hayashi}},
  \bibinfo{author}{\bibfnamefont{M.}~\bibnamefont{Tanaka}}, \bibnamefont{and}
  \bibinfo{author}{\bibfnamefont{A.}~\bibnamefont{Asamitsu}},
  \bibinfo{journal}{J. Appl. Phys.} \textbf{\bibinfo{volume}{87}},
  \bibinfo{pages}{4673} (\bibinfo{year}{2000}).

\bibitem[{\citenamefont{Bruno and Wunderlich}(1998)}]{BrWu98}
\bibinfo{author}{\bibfnamefont{P.}~\bibnamefont{Bruno}} \bibnamefont{and}
  \bibinfo{author}{\bibfnamefont{J.}~\bibnamefont{Wunderlich}},
  \bibinfo{journal}{J. Appl. Phys.} \textbf{\bibinfo{volume}{84}},
  \bibinfo{pages}{978} (\bibinfo{year}{1998}).

\bibitem[{\citenamefont{Vurgaftman and Meyer}(2003)}]{VuMe03}
\bibinfo{author}{\bibfnamefont{I.}~\bibnamefont{Vurgaftman}} \bibnamefont{and}
  \bibinfo{author}{\bibfnamefont{J.~R.} \bibnamefont{Meyer}},
  \bibinfo{journal}{Phys. Rev. B} \textbf{\bibinfo{volume}{67}},
  \bibinfo{pages}{125209} (\bibinfo{year}{2003}).

\bibitem[{\citenamefont{Voskoboynikov et~al.}(2000)\citenamefont{Voskoboynikov,
  Lin, Lee, and Tretyak}}]{VoShLe00}
\bibinfo{author}{\bibfnamefont{A.}~\bibnamefont{Voskoboynikov}},
  \bibinfo{author}{\bibfnamefont{S.~S.} \bibnamefont{Lin}},
  \bibinfo{author}{\bibfnamefont{C.~P.} \bibnamefont{Lee}}, \bibnamefont{and}
  \bibinfo{author}{\bibfnamefont{O.}~\bibnamefont{Tretyak}},
  \bibinfo{journal}{J. Appl. Phys.} \textbf{\bibinfo{volume}{87}},
  \bibinfo{pages}{387} (\bibinfo{year}{2000}).

\bibitem[{\citenamefont{Gruber et~al.}(2001)\citenamefont{Gruber, Keim,
  Fiederling, Reuscher, Ossau, Schmidt, Molenkamp, and Waag}}]{GrKeFi01}
\bibinfo{author}{\bibfnamefont{T.}~\bibnamefont{Gruber}},
  \bibinfo{author}{\bibfnamefont{M.}~\bibnamefont{Keim}},
  \bibinfo{author}{\bibfnamefont{R.}~\bibnamefont{Fiederling}},
  \bibinfo{author}{\bibfnamefont{G.}~\bibnamefont{Reuscher}},
  \bibinfo{author}{\bibfnamefont{W.}~\bibnamefont{Ossau}},
  \bibinfo{author}{\bibfnamefont{G.}~\bibnamefont{Schmidt}},
  \bibinfo{author}{\bibfnamefont{L.~W.} \bibnamefont{Molenkamp}},
  \bibnamefont{and} \bibinfo{author}{\bibfnamefont{A.}~\bibnamefont{Waag}},
  \bibinfo{journal}{Appl. Phys. Lett.} \textbf{\bibinfo{volume}{78}},
  \bibinfo{pages}{1101} (\bibinfo{year}{2001}).

\bibitem[{\citenamefont{Beletskii et~al.}(2005)\citenamefont{Beletskii, Berman,
  and Borysenko}}]{BeBeBo05}
\bibinfo{author}{\bibfnamefont{N.~N.} \bibnamefont{Beletskii}},
  \bibinfo{author}{\bibfnamefont{G.~P.} \bibnamefont{Berman}},
  \bibnamefont{and} \bibinfo{author}{\bibfnamefont{S.~A.}
  \bibnamefont{Borysenko}}, \bibinfo{journal}{Phys. Rev. B}
  \textbf{\bibinfo{volume}{71}}, \bibinfo{pages}{125325}
  (\bibinfo{year}{2005}).

\bibitem[{\citenamefont{Zohta et~al.}(1989)\citenamefont{Zohta, Nozu, and
  Obara}}]{ZoNoOb89}
\bibinfo{author}{\bibfnamefont{Y.}~\bibnamefont{Zohta}},
  \bibinfo{author}{\bibfnamefont{T.}~\bibnamefont{Nozu}}, \bibnamefont{and}
  \bibinfo{author}{\bibfnamefont{M.}~\bibnamefont{Obara}},
  \bibinfo{journal}{Phys. Rev. B} \textbf{\bibinfo{volume}{39}},
  \bibinfo{pages}{1375} (\bibinfo{year}{1989}).

\bibitem[{\citenamefont{Palevski et~al.}(1990)\citenamefont{Palevski, Beltram,
  Capasso, Pfeiffer, and West}}]{PaBeCa90}
\bibinfo{author}{\bibfnamefont{A.}~\bibnamefont{Palevski}},
  \bibinfo{author}{\bibfnamefont{F.}~\bibnamefont{Beltram}},
  \bibinfo{author}{\bibfnamefont{F.}~\bibnamefont{Capasso}},
  \bibinfo{author}{\bibfnamefont{L.}~\bibnamefont{Pfeiffer}}, \bibnamefont{and}
  \bibinfo{author}{\bibfnamefont{K.~W.} \bibnamefont{West}},
  \bibinfo{journal}{Phys. Rev. Lett.} \textbf{\bibinfo{volume}{65}},
  \bibinfo{pages}{1929} (\bibinfo{year}{1990}).

\bibitem[{\citenamefont{Macks et~al.}(1996)\citenamefont{Macks, Brown, Clark,
  Starrett, Reed, Deshpande, Fernando, and Frensley}}]{MaBrCl96}
\bibinfo{author}{\bibfnamefont{L.~D.} \bibnamefont{Macks}},
  \bibinfo{author}{\bibfnamefont{S.~A.} \bibnamefont{Brown}},
  \bibinfo{author}{\bibfnamefont{R.~G.} \bibnamefont{Clark}},
  \bibinfo{author}{\bibfnamefont{R.~P.} \bibnamefont{Starrett}},
  \bibinfo{author}{\bibfnamefont{M.~A.} \bibnamefont{Reed}},
  \bibinfo{author}{\bibfnamefont{M.~R.} \bibnamefont{Deshpande}},
  \bibinfo{author}{\bibfnamefont{C.~J.~L.} \bibnamefont{Fernando}},
  \bibnamefont{and} \bibinfo{author}{\bibfnamefont{W.~R.}
  \bibnamefont{Frensley}}, \bibinfo{journal}{Phys. Rev. B}
  \textbf{\bibinfo{volume}{54}}, \bibinfo{pages}{4857} (\bibinfo{year}{1996}).

\bibitem[{\citenamefont{Newaz et~al.}(2005)\citenamefont{Newaz, Song, and
  Mendez}}]{NeSoMe05}
\bibinfo{author}{\bibfnamefont{A.}~\bibnamefont{Newaz}},
  \bibinfo{author}{\bibfnamefont{W.}~\bibnamefont{Song}}, \bibnamefont{and}
  \bibinfo{author}{\bibfnamefont{E.~E.} \bibnamefont{Mendez}},
  \bibinfo{journal}{Phys. Rev. B} \textbf{\bibinfo{volume}{71}},
  \bibinfo{pages}{195303} (\bibinfo{year}{2005}).

\bibitem[{\citenamefont{Furdyna}(1988)}]{Fu88}
\bibinfo{author}{\bibfnamefont{J.~K.} \bibnamefont{Furdyna}},
  \bibinfo{journal}{J. Appl. Phys.} \textbf{\bibinfo{volume}{64}},
  \bibinfo{pages}{R29} (\bibinfo{year}{1988}).

\bibitem[{\citenamefont{Oiwa et~al.}(2004)\citenamefont{Oiwa, Moriya,
  Kashimura, and Munekata}}]{OiMoKa04}
\bibinfo{author}{\bibfnamefont{A.}~\bibnamefont{Oiwa}},
  \bibinfo{author}{\bibfnamefont{R.}~\bibnamefont{Moriya}},
  \bibinfo{author}{\bibfnamefont{Y.}~\bibnamefont{Kashimura}},
  \bibnamefont{and} \bibinfo{author}{\bibfnamefont{H.}~\bibnamefont{Munekata}},
  \bibinfo{journal}{J. Magn. Magn. Mater.} \textbf{\bibinfo{volume}{272}}
  (\bibinfo{year}{2004}).

\bibitem[{\citenamefont{Furdyna et~al.}(2004)\citenamefont{Furdyna, Wojtowicz,
  Liu, Yu, Walukiewicz, Vurgaftman, and Meyer}}]{FuWoLi04}
\bibinfo{author}{\bibfnamefont{J.}~\bibnamefont{Furdyna}},
  \bibinfo{author}{\bibfnamefont{T.}~\bibnamefont{Wojtowicz}},
  \bibinfo{author}{\bibfnamefont{X.}~\bibnamefont{Liu}},
  \bibinfo{author}{\bibfnamefont{K.~M.} \bibnamefont{Yu}},
  \bibinfo{author}{\bibfnamefont{W.}~\bibnamefont{Walukiewicz}},
  \bibinfo{author}{\bibfnamefont{I.}~\bibnamefont{Vurgaftman}},
  \bibnamefont{and} \bibinfo{author}{\bibfnamefont{J.~R.} \bibnamefont{Meyer}},
  \bibinfo{journal}{J. Phys.: Condens. Matter} \textbf{\bibinfo{volume}{16}},
  \bibinfo{pages}{S5499} (\bibinfo{year}{2004}).

\bibitem[{\citenamefont{Dietl}(2002)}]{Di02}
\bibinfo{author}{\bibfnamefont{T.}~\bibnamefont{Dietl}},
  \bibinfo{journal}{Semicond. Sci. Technol.} \textbf{\bibinfo{volume}{17}},
  \bibinfo{pages}{377} (\bibinfo{year}{2002}).

\bibitem[{\citenamefont{Havu et~al.}(2005)\citenamefont{Havu, Tuomisto,
  V\"a\"an\"anen, Puska, and Nieminen}}]{HaTuVa05}
\bibinfo{author}{\bibfnamefont{P.}~\bibnamefont{Havu}},
  \bibinfo{author}{\bibfnamefont{N.}~\bibnamefont{Tuomisto}},
  \bibinfo{author}{\bibfnamefont{R.}~\bibnamefont{V\"a\"an\"anen}},
  \bibinfo{author}{\bibfnamefont{M.~J.} \bibnamefont{Puska}}, \bibnamefont{and}
  \bibinfo{author}{\bibfnamefont{R.~M.} \bibnamefont{Nieminen}},
  \bibinfo{journal}{Phys. Rev. B} \textbf{\bibinfo{volume}{71}},
  \bibinfo{pages}{235301} (\bibinfo{year}{2005}).

\bibitem[{\citenamefont{Vassell et~al.}(1983)\citenamefont{Vassell, Lee, and
  Lockwood}}]{VaLeLo83}
\bibinfo{author}{\bibfnamefont{M.~O.} \bibnamefont{Vassell}},
  \bibinfo{author}{\bibfnamefont{J.}~\bibnamefont{Lee}}, \bibnamefont{and}
  \bibinfo{author}{\bibfnamefont{H.~F.} \bibnamefont{Lockwood}},
  \bibinfo{journal}{J. Appl. Phys.} \textbf{\bibinfo{volume}{54}},
  \bibinfo{pages}{5206} (\bibinfo{year}{1983}).

\bibitem[{\citenamefont{Cahay et~al.}(1987)\citenamefont{Cahay, McLennan,
  Datta, and Lundstrom}}]{CaMcDa87}
\bibinfo{author}{\bibfnamefont{M.}~\bibnamefont{Cahay}},
  \bibinfo{author}{\bibfnamefont{M.}~\bibnamefont{McLennan}},
  \bibinfo{author}{\bibfnamefont{S.}~\bibnamefont{Datta}}, \bibnamefont{and}
  \bibinfo{author}{\bibfnamefont{M.~S.} \bibnamefont{Lundstrom}},
  \bibinfo{journal}{Appl. Phys. Lett.} \textbf{\bibinfo{volume}{50}},
  \bibinfo{pages}{612} (\bibinfo{year}{1987}).

\bibitem[{\citenamefont{P\"otz}(1989)}]{Po89}
\bibinfo{author}{\bibfnamefont{W.}~\bibnamefont{P\"otz}}, \bibinfo{journal}{J.
  Appl. Phys.} \textbf{\bibinfo{volume}{66}}, \bibinfo{pages}{2458}
  (\bibinfo{year}{1989}).

\bibitem[{\citenamefont{Tsu and Esaki}(1973)}]{TsEs73}
\bibinfo{author}{\bibfnamefont{R.}~\bibnamefont{Tsu}} \bibnamefont{and}
  \bibinfo{author}{\bibfnamefont{L.}~\bibnamefont{Esaki}},
  \bibinfo{journal}{Appl. Phys. Lett.} \textbf{\bibinfo{volume}{22}},
  \bibinfo{pages}{562} (\bibinfo{year}{1973}).

\bibitem[{\citenamefont{Berk et~al.}(1994)\citenamefont{Berk, Kamenev,
  Palevski, Pfeiffer, and West}}]{BeKaPa94}
\bibinfo{author}{\bibfnamefont{Y.}~\bibnamefont{Berk}},
  \bibinfo{author}{\bibfnamefont{A.}~\bibnamefont{Kamenev}},
  \bibinfo{author}{\bibfnamefont{A.}~\bibnamefont{Palevski}},
  \bibinfo{author}{\bibfnamefont{L.~N.} \bibnamefont{Pfeiffer}},
  \bibnamefont{and} \bibinfo{author}{\bibfnamefont{K.~W.} \bibnamefont{West}},
  \bibinfo{journal}{Phys. Rev. B} \textbf{\bibinfo{volume}{50}},
  \bibinfo{pages}{15420} (\bibinfo{year}{1994}).

\end{thebibliography}
\end{document}